\documentclass[aps,prb,twocolumn,amsmath,amssymb,longbibliography,noeprint,maxbibnames=2]{revtex4-1}

\usepackage[%
  colorlinks=true,
  urlcolor=blue,
  linkcolor=blue,
  citecolor=blue
]{hyperref}

\usepackage{graphicx}
\usepackage{dcolumn}
\usepackage{bm}
\usepackage{ulem}
\usepackage{braket}
\usepackage{color}

\begin{document}

\title{Generalization of the Franck-Condon model for phonon excitations by resonant inelastic X-ray scattering }
\author{Andrey Geondzhian$^{1}$ and Keith Gilmore$^{2}$ }
\affiliation{$^1$European Synchrotron Radiation Facility, 71 avenue des Martyrs, 38043 Grenoble, France \\
$^2$Condensed Matter Physics $\&$ Materials Science Division,
Brookhaven National Laboratory, Upton, NY 11973-5000, USA}
\date{\today}
\begin{abstract}
    Resonant inelastic X-ray scattering (RIXS) is increasingly used to quantify vibronic interactions in materials.  In the case of periodic systems, this is most often done through fitting experimental results to a parameterized, but exact analytical solution of a simple Holstein Hamiltonian that consists of a single electronic level coupled linearly to a single Einstein vibrational mode.  Working within this standard framework, we consider the impact of minor generalizations of this model, namely, introducing a second Einstein oscillator, and allowing the curvature of the excited-state potential energy surface to differ from that of the ground-state potential energy surface.  We find that dynamics occurring in the RIXS intermediate (excited) state considerably alter the quantitative interpretation of the spectral features observed in the RIXS final state.  This complicates the use of the single mode model when multiple phonon modes are active.  Our generalized model may in principle by substituted in this case, though we find that accurate quantitative results rely on knowledge of the excited-state potential energy surface, though this typically is not known.
\end{abstract}
\maketitle
\section{Introduction}

Strongly correlated electron materials exhibit intricate phase diagrams and profound responses to changes in control parameters.  This makes these materials difficult to understand, but at the same time confers upon them great technological promise.  While this behavior may be viewed as originating from multiple competing orders, it is advantageous to instead view these materials as having intertwined orders, stemming as much from the cooperation as the competition between various degrees of freedom \cite{Tranquada.RMP}.  For example, it is the cooperation between strong spin-orbit coupling and the otherwise insufficient Coulomb interaction that opens a Mott gap in several $5d$ oxides \cite{Kim.2008, Jackeli.2009}.  Clearly, improving our understanding of strongly correlated materials will require a better quantification of the interactions between the charge, spin, orbital and lattice subsystems.  In particular, the variation of these coupling strengths across the Brillouin zone is likely to have considerable impact.

Taking superconductivity as an example, while the interaction between the charge and lattice degrees of freedom is responsible for Cooper pairing in conventional BCS superconductors, it is insufficient to explain unconventional superconductivity.  Nevertheless, electron-phonon coupling may still play an assisting role to a more fundamental pairing mechanism in unconventional superconductors \cite{Johnston.2010}.  For anisotropic unconventional superconductivity, the momentum dependence of the electron-phonon coupling strength may be of considerable importance \cite{Andersen.1996, Bulut.1996} even though it does not significantly affect Cooper pairing in isotropic conventional BCS superconductors.  For this and other reasons, it is of considerable interest to ascertain the variation in the electron-phonon coupling strength throughout the Brillouin zone.  However, since for most physical phenomena the electron-phonon interaction manifests itself as an average over the Brillouin zone experimental access to the momentum dependence of the electron-phonon interaction strength is limited.

The momentum dependence of the electron-lattice interaction can be measured, to an extent, by techniques that map the phonon dispersions, such as inelastic neutron scattering \cite{Pintschovius.2005} and non-resonant inelastic X-ray scattering \cite{Shukla.2003}.  In these cases, the electron-phonon interaction produces a finite linewidth to the phonon peaks.  However, in regions of strong coupling, the linewidth becomes sufficiently broad to limit the accuracy of such quantification.  Angle resolved photoemission \cite{Cuk.2005} and scanning tunneling spectroscopy \cite{Schrieffer.1963} alternatively can probe electron-phonon coupling with control over the electron momentum, but yield values averaged over the phonon wavevector.  Optical techniques, such as Raman scattering \cite{Ferrari.2007}, are surface sensitive and limited to the zone center.

Resonant inelastic x-ray scattering (RIXS) is increasingly used as an alternative method to quantify electron-lattice coupling.  With few exceptions, most analyses of experimental data from periodic systems employ a fitting procedure predicated on a Holstein model with a single electronic level coupled to a single Einstein vibration \cite{Ament2011}.  Such a model has historically been used to describe vibronic interactions in molecules measured by optical Raman spectroscopy \cite{Hoffmann.2002}.  In the canonical case, one observes a series of loss features appearing at integer multiples of the phonon energy with decreasing peak height. Within this model, the RIXS intensity of the $n$th phonon harmonic in the loss spectrum is given by the square of
\begin{equation}
    \label{eq:simple.FC}
    A_{n}=\sum_m\frac{B_{n,m}(g)B_{m,0}(g)}{z-\omega_0(m-g) } \ ,
\end{equation}
where the summation runs over all intermediate state phonon occupancies $m$, $\omega_0$ is the vibrational energy, $g$ is the electron-phonon coupling strength, and $z=\Delta+i\gamma/2$ with $\Delta$ being the detuning between the energy of the electronic excitation and the incident photon energy, and $\gamma$ the inverse lifetime of the core-hole.  $B(g)$ is a Franck-Condon factor and the use of $B_{n,m}(g)$ is shorthand for  $B_{max(n,m),min(n,m)}(g)$.

Equation \eqref{eq:simple.FC} has been used to quantify the vibronic coupling strength in several materials.  Probing the Ti L$_3$ edge, Fatale {\it et al}.~found that the 65 meV TO mode of BaTiO$_3$ had a coupling parameter of $g \approx 19$ \cite{Fatale.2016} while Moser {\it et al}.~inferred $g=1.9$ for the 95 meV LO mode of anatase TiO$_2$ \cite{Moser.2015}.  At the O-K edge, Meyers {\it et al}.~extracted a coupling of $g=14$ for the 105 meV O-Ti LO$_4$ mode in multilayers of SrIrO$_3$ / SrTiO$_3$ \cite{Meyers.2018} and Vale {\it et al}.~found a similar value in the 5$d$ system $\alpha$-Li$_2$IrO$_3$ for the oxygen $A_u$ mode at 70 meV \cite{Vale.2019}.  For NdBa$_2$Cu$_3$O$_6$, Rossi {\it et al}.~obtained a value of $g\approx 2$ at low momenta transfer for the 70 meV Cu-O bond stretching mode \cite{Rossi.2019}.

To quantitatively analyze vibronic coupling in from RIXS measurements on periodic materials it will be desirable in future work to adopt a more sophisticated description of phonon excitations than that provided by the Holstein Hamiltonian underlying Eq.~\eqref{eq:simple.FC}.  This will likely start from a Fr\"{o}hlich-type Hamiltonian with a fully momentum dependent electron-phonon coupling parameter including the momenta of both the phonons and electrons.  Devereaux {\it et al.} have made initial efforts in this direction \cite{Devereaux.Frohlich}, but such work is thus far limited to the single phonon contribution with further extension appearing to present numerical challenges.  Since it is likely that Eq.~\eqref{eq:simple.FC} will continue to be used to analyze RIXS data it is worthwhile to probe the behavior of this model more deeply.  

Braicovich {\it et al}.~have recently presented an in depth discussion on the nature of Eq.~\eqref{eq:simple.FC} for the conceptually ideal model of a single local electronic level coupled linearly to a single local Einstein mode \cite{Braicovich.2019}.  Here, we consider the impact of making two simple generalizations to the basic model that are both routinely encountered and have significant impact on the interpretation of the data.  First, in Section II, we consider the situation of a single electronic level coupled to two vibrational modes.  The temptation is to sum the two contributions independently according to Eq.~\eqref{eq:simple.FC}.  We find this to be a poor approximation for typical coupling strengths.  This is due to the fact that the modes may mix via the electronic excitation, shifting spectral weight to regions of higher energy loss.  Also, even when the final state consists of only a single quantum in a single mode, intermediate states with high occupancy of both modes can contribute significantly to this final state.  This further reveals that the intensity of each feature will depend on the coupling strengths of both modes.

Second, a basic assumption of Eq.~\eqref{eq:simple.FC} is that the vibrational frequency is the same in the ground state and the excited state.  In Section III, we investigate the consequences of relaxing this assumption.  We find that changes in the potential energy surface between the ground and excited state lead to clear variations in the phonon peak intensities observed in the final spectrum.  This effect becomes important when the RIXS intermediate state excites localized vibrations and either the symmetry or spatial variation of the excited-state charge density differs appreciably from the ground-state charge density.  For example, this could be due to an excited-state Jahn-Teller distortion and be viewed as a transient, excited-state polaron with a local potential that differs from that of the corresponding ground-state polaron.  This implies that to properly extract a coupling constant from the final RIXS spectrum one must know the intermediate state potential energy surface.

There is a fundamental question as to the relation between the coupling value inferred from a RIXS measurement and the transport electron-phonon coupling parameter.  Phonons are excited by the altered potential of the RIXS intermediate state, which contains both an excited electron and a core-hole -- an excitonic configuration \cite{Geondzhian.2018}.  While the potential of an added electron and an exciton may be approximately equivalent in a certain limit, they will generally differ.  It remains an open question to quantitatively determine the relation between the effective coupling constant measured by RIXS and the near-equilibrium transport electron-phonon coupling parameter.  We do not address this point in the present work, but instead consider and refer to a generic vibronic coupling between the electronic and lattice systems without specifying any relation to the transport electron-phonon parameter.

\section{RIXS signal with two vibrational modes}
\label{sec:2modes}

The amplitude for the phonon contribution to the RIXS signal expressed in Eq.~\eqref{eq:simple.FC} originates from considering the basic Holstein Hamiltonian for a single electronic level coupled linearly to a single Einstein mode
~
\begin{equation}
\label{eq:hh1}
    H=\epsilon_0 d^+d+\omega_0 b^+ b + M d^+d (b+b^+)  \
\end{equation}

\noindent where the dimensionful coupling constant, $M$, relates to the dimensionless coupling parameter, $g$, by $g=(M/\omega_0)^2$.  Due to the localized nature of the electronic level and simple linear coupling, it is possible to re-express Eq.~\eqref{eq:hh1} in diagonal form as a displaced harmonic oscillator and obtain an exact solution by applying a Lang-Firsov canonical transformation ($\bar{H}=e^{S}He^{-S}$) with the generating function $S= \sqrt{g}d^+d(b^+-b)$ \cite{Lang.1963}.  Overlap of the resulting excited-state vibrational wavefunctions with the ground-state vibrational wavefunctions gives the Franck-Condon factors in Eq.~\eqref{eq:simple.FC}.

The corresponding Holstein Hamiltonian with a single electronic level coupled to two Einstein modes that are otherwise independent is
~
\begin{equation}
\label{eq:hh2}
    H=\epsilon_0 d^+d+\sum_{\lambda=1,2} \omega_\lambda b^+_\lambda b_\lambda + \sum_{\lambda=1,2} M_\lambda d^+d (b_\lambda+b_\lambda^+)  \ .
\end{equation}

\noindent By analogy to the single oscillator, we can cast this Hamiltonian into the form of a two dimensional displaced harmonic oscillator and again obtain an exact solution.  In this case, the Lang-Firsov generating function for the canonical transformation is
$S= \sum_\lambda \sqrt{g_\lambda}d^+d(b_\lambda^+-b_\lambda)$.
The generalization of Eq.~\eqref{eq:simple.FC} for the RIXS phonon intensity is 
~
\begin{equation}
    \label{eq:general.FC}
    A_{n_{1}n_{2}}=\sum_{m_1,m_2}\frac{D_{n_{1}m_{1}}^{n_{2}m_{2}}(g_{1},g_{2})D_{m_{1}0}^{m_{2}0}(g_{1},g_{2})}{z-\sum_{\lambda=1,2} \omega_\lambda(m_\lambda-g_\lambda ) } \ ,
\end{equation}

\noindent where 
~
\begin{eqnarray*}
D_{n_{1}m_{1}}^{n_{2}m_{2}}(g_{1},g_{2}) &=& B_{max(n_1,m_1), min(n_1,m_1)}(g_1) \\
 &\times& B_{max(n_2,m_2),min(n_2,m_2)}(g_2)
\end{eqnarray*}

\noindent is a product of Franck-Condon factors.  This expression may be further generalized to higher dimensions.  Although this result is obtained in close analogy to the one dimensional case there are important impacts on the behavior of the RIXS intensities that we now consider.

\begin{figure}
\includegraphics[width=1.0\linewidth,angle=0]{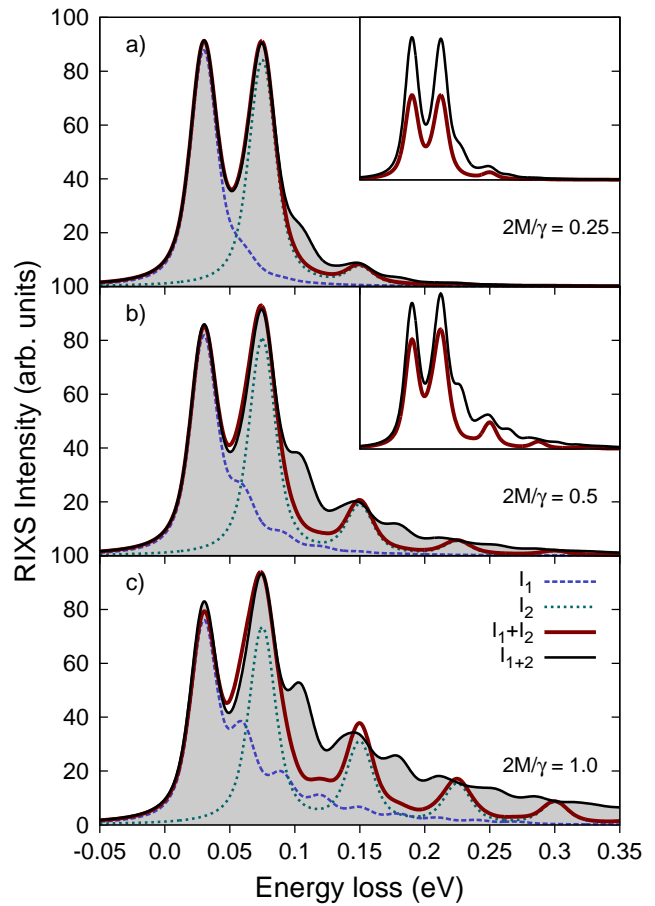}
    \caption{RIXS loss profile for two equally coupled modes. The contribution from the two mode model that includes mixing of the modes (black curve an grey shading) is compared to the independent mode approximation (red curve).  In the latter case, the individual contributions from the two modes are given by the dashed blue and dotted green curves. In each panel, the energies of the two modes are $\omega_1=30$ meV and $\omega_2=75$ meV, the core-hole lifetime is $\gamma/2=200$ meV and the excitation was tuned to the resonance.  The dimensionful coupling strength of the two modes are set equal in each panel and are $2M/\gamma=0.25$ (a), $2M/\gamma=0.5$ (b), and $2M/\gamma=1$ (c).  The elastic lines are not shown.}
    \label{fig:equal}
\end{figure}{}

\begin{figure}
\includegraphics[width=0.7\linewidth,angle=270]{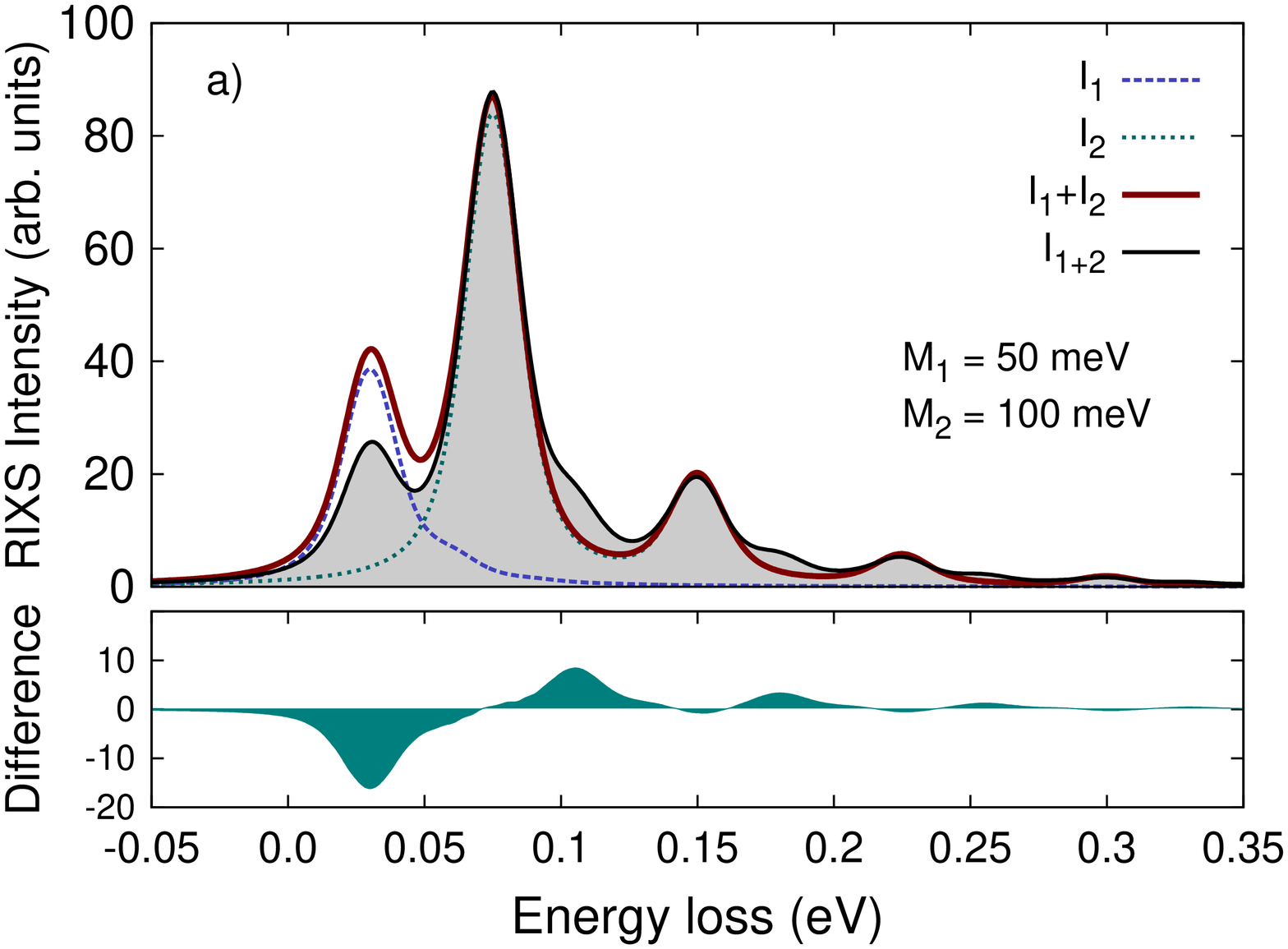}
\includegraphics[width=0.7\linewidth,angle=270]{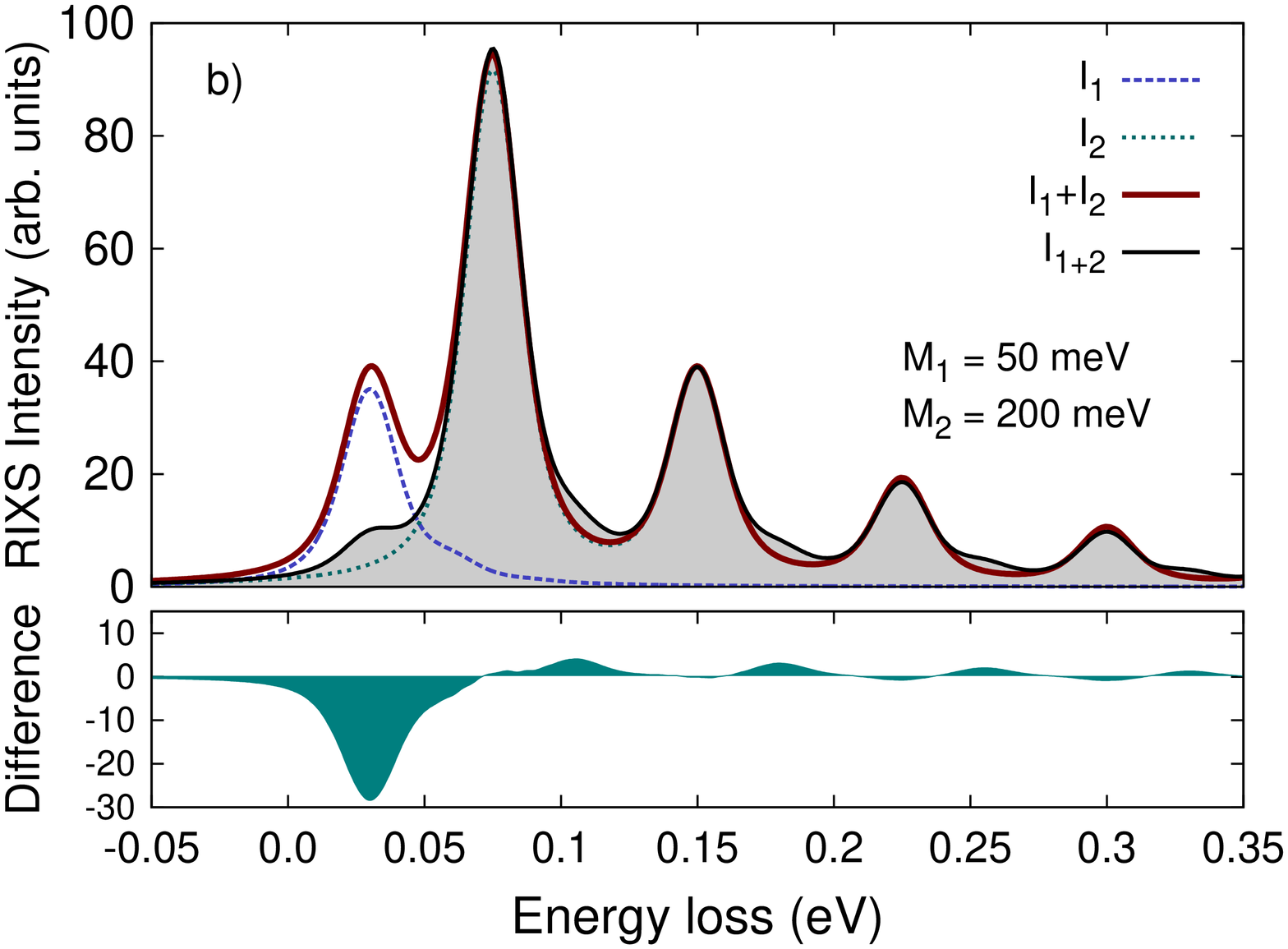}
    \caption{RIXS loss profile for two unequally coupled modes.  The color scheme and parameters are the same as in Fig.~\ref{fig:equal} except that $M_1$ is fixed at 50 meV and $M_2$ is either 100 meV (a) or 200 meV (b).  The intensity difference between the independent mode approximation and the two mode model is given below each RIXS profile.}
    \label{fig:high}
\end{figure}

\begin{figure}
\includegraphics[width=0.6\linewidth,angle=270]{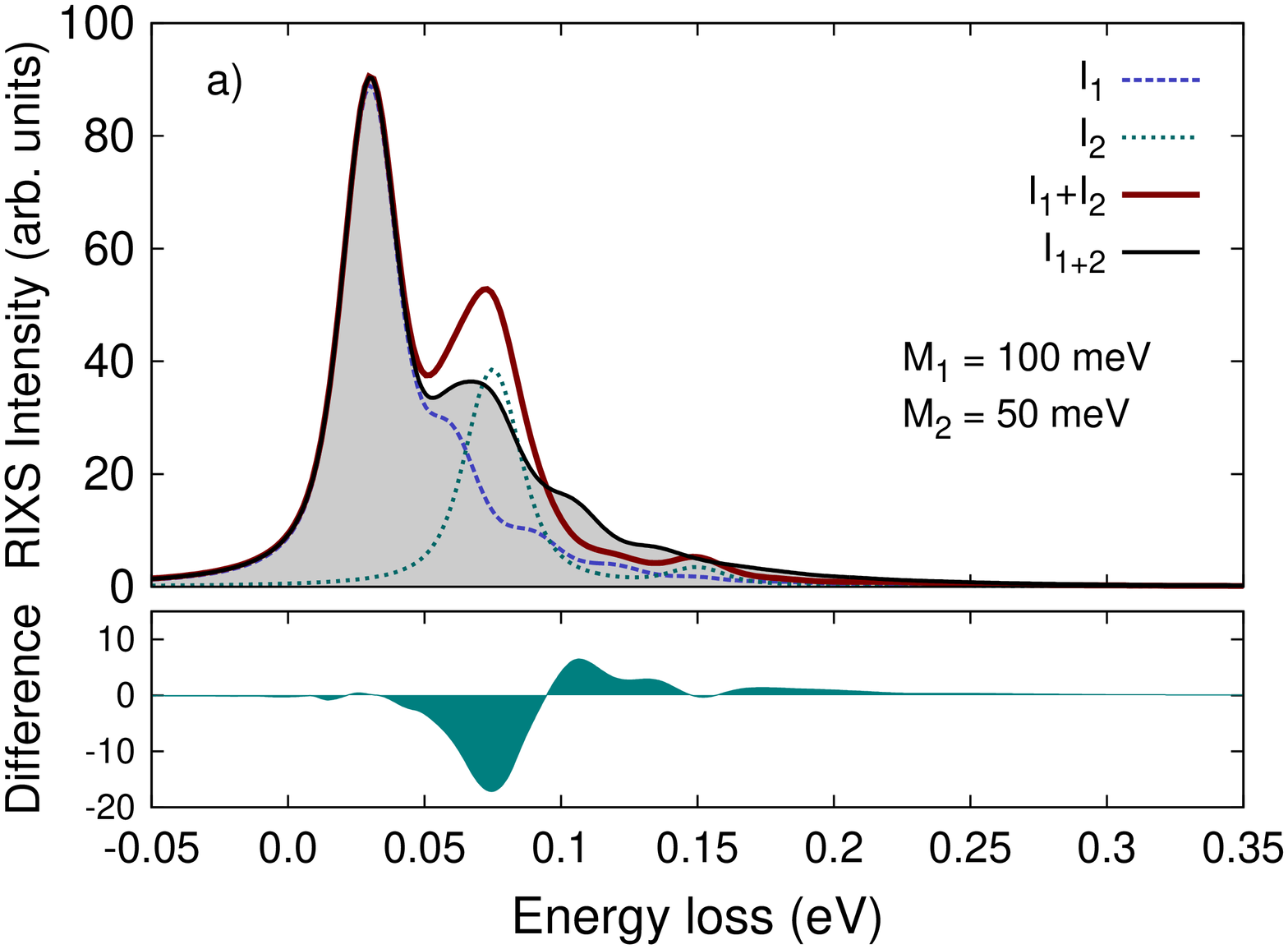}
\includegraphics[width=0.6\linewidth,angle=270]{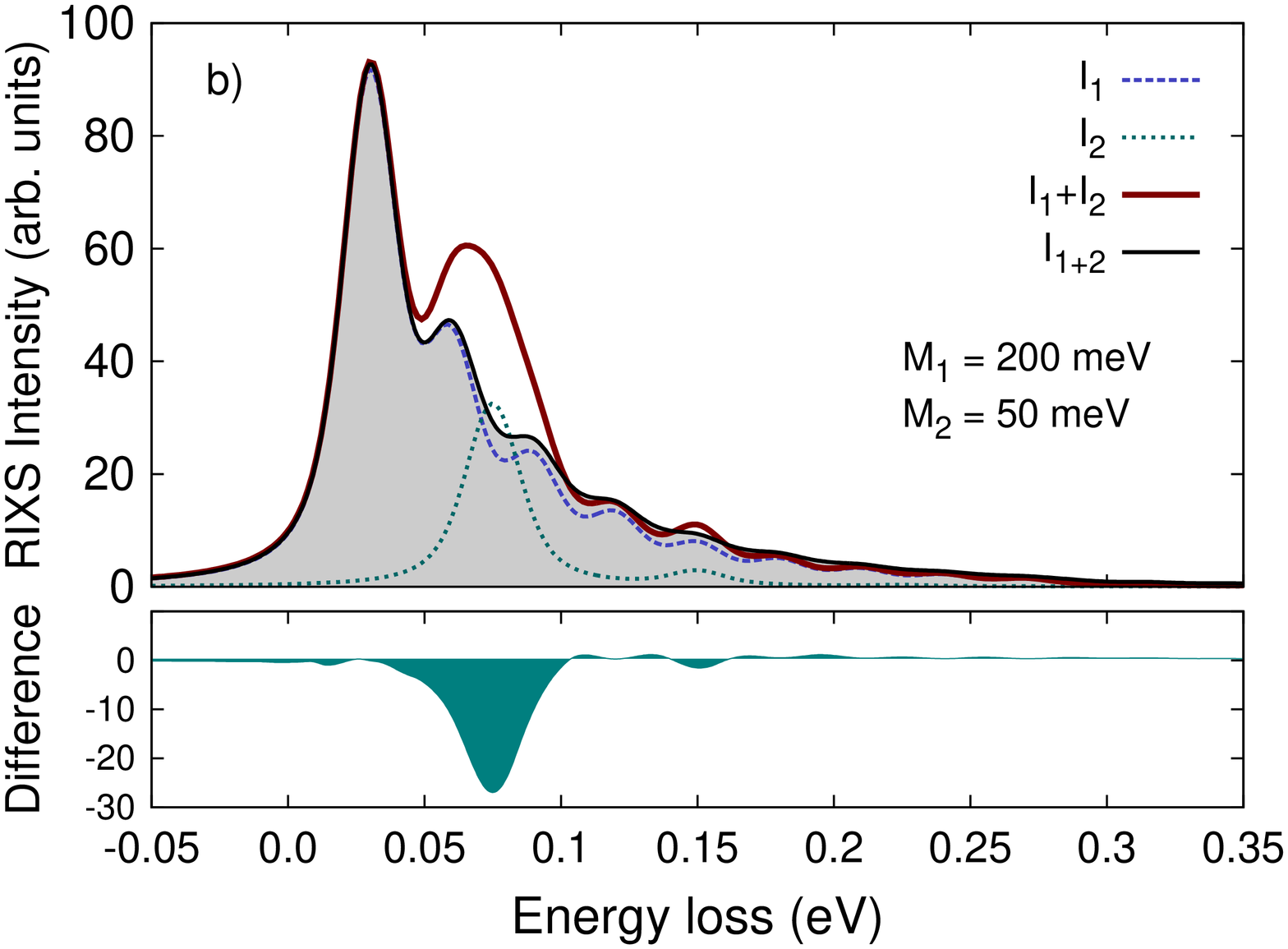}
    \caption{RIXS loss profile for two unequally coupled modes.  The color scheme and parameters are the same as in Fig.~\ref{fig:equal} except that $M_1$ is either 100 meV (a) or 200 meV (b) while $M_2$ is fixed at 50 meV.  The intensity difference between the independent mode approximation and the two mode model is given below each RIXS profile.}
    \label{fig:low}
\end{figure}

To illustrate the differences between the usual single mode model of Eq.~\eqref{eq:simple.FC} and the two mode model in Eq.~\eqref{eq:general.FC} we consider a low energy mode with energy $\omega_1=30$ meV and a high energy mode of $\omega_2=75$ meV.  The two modes have coupling parameters $M_1$ ($g_1$) and $M_2$ ($g_2$), respectively.  Rather than classifying the coupling strength with $g=(M/\omega)^2$ we instead use $2M/\gamma$ as the dimensionless measure of the coupling strength.  It is important to note that while $\gamma$ gives the full width at half maximum of the Lorentizian contribution to the absorption or photoemission lineshape, the denominators of Eqs.~\eqref{eq:simple.FC} and \eqref{eq:general.FC} contain $z=\Delta+i\gamma/2$, that is, the half width at half maximum.  This point has caused some confusion in the quantitative application of Eq.~\eqref{eq:simple.FC} to experimental data.  We take the core-hole lifetime broadening to be $\gamma=400$ meV throughout ($\gamma/2=200$ meV), which is intermediate between typical values for the oxygen 1s level and the copper 2p level.  In Figs.~\ref{fig:equal}-\ref{fig:low} we compare the RIXS profile from Eq.~\eqref{eq:general.FC}, which we refer to as the two mode model, to that from Eq.~\eqref{eq:simple.FC}, which we call the independent mode approximation.  Specifically, the independent mode approximation consists of the signal constructed by assuming that the two phonon modes contribute independently to the RIXS profile and summing their intensities using the single mode model in Eq.~\eqref{eq:simple.FC} separately for each mode.

Figure \ref{fig:equal} considers the special case that $M_1=M_2=M$ varying $M$ from weak ($2M/\gamma=0.25$) to moderate ($2M/\gamma=1$).  For each value of $2M/\gamma$ we compare the RIXS signal obtained from the two mode model to the independent mode approximation.  Within each panel, the independent mode contributions are given by the dashed blue (30 meV mode) and dotted green (75 meV mode) lines, their summed contributions are indicated by the red curve, and the result of the two mode model is given by the black curve with grey shading.  The elastic line has been removed for clarity.

A few points warrant mention.  The two mode model converges to the independent mode approximation in the weak coupling limit as expected (Fig.~\ref{fig:equal}a), although a small difference between the two results is still evident in the vicinity of $\omega_1+\omega_2 = 105$ meV.  This occurs because the two mode model allows final states with both oscillators simultaneously occupied, {\it e.g.} $\ket{n_1,n_2}=\ket{1,1}$, whereas such a final state is not accessible within the independent mode approximation.  The importance of these mixed final states within the two mode model, containing excitations of both oscillators, becomes more pronounced as the coupling strength increases (Figs.~\ref{fig:equal}b-c).  While the intensities around the first harmonic of each mode remain essentially unchanged, the two mode model develops a significant high energy tail that is attenuated in the independent mode approximation.  To match the result from the two mode model by independently summing the single modes it would be necessary to erroneously increase the value of the coupling strengths in order to account for the greater intensity in the high energy loss region.  

The agreement in intensities at the first harmonics occurs only because we have scaled the overall spectra in this way.  The insets of Figs.~\ref{fig:equal}a-b present the properly normalized spectra.  In all cases, the two mode model has a greater total intensity summed over the phonon contribution.  Consequently, the elastic line (not shown) is reduced in the two mode model with respect to the independent mode approximation.  This indicates that the two mode model has a larger effective coupling strength.  Since the elastic line is not generally used as an intensity reference the scaling of the spectra may not seem important.  However, proper scaling must be enforced when analyzing phonon satellites to other features, such as $d$-$d$ excitations.

{}

Figure \ref{fig:high} demonstrates the effect of adding a weakly coupled low energy mode ($M_1=50$ meV, $2M_1/\gamma=0.25$) to a more strongly coupled high energy mode.  We consider overall weak coupling ($2M_2/\gamma=0.5$, $M_2/M_1=2$) in Fig.~\ref{fig:high}a and a more strongly coupled high energy mode ($2M_2/\gamma=1$, $M_2/M_1=4$) in Fig.~\ref{fig:high}b.  The largest difference between the two mode model and the independent mode approximation occurs below the energy of the first harmonic of the high energy mode; the high energy mode progression is barely affected.  Such a scenario has likely been observed recently in $\alpha$-Li$_2$IrO$_3$ \cite{Vale.2019}.  Below each set of spectra we plot the difference between the independent mode approximation and the two mode model.  The difference plot clarifies that the significant loss of spectral weight at the first harmonic of the lower energy mode is redistributed as a satellite feature to each harmonic of the higher energy mode.  In particular, the positive regions of the difference plot are spaced in energy by $\omega_2=75$ meV and are shifted by $\omega_1=30$ meV from the harmonics of the second mode.  This may essentially be viewed as a convolution of the single mode spectrum for the low energy mode with the harmonics of the high energy mode, and a concomitant reduction of the low energy mode satellites associated with the elastic line.  However, the actual effect is more subtle when the spectra are properly normalized.

The degree of attenuation of the first harmonic of the low energy mode increases with $M_2/M_1$.  Fitting experimental data that resemble the two mode result (black curve) with the independent mode approximation would yield a coupling constant for the low energy mode much smaller than its actual value.  On the other hand, a reasonable result for the high energy mode could be obtained with the independent mode summation in this case.  In general, it appears that if a high energy mode has a much larger dimensionful coupling parameter than all other modes the contributions of the weakly coupled modes to the RIXS loss profile are effectively attenuated, leaving the contribution of the strongly coupled high energy mode largely isolated.  More precisely, the spectral weight of the weakly coupled low energy modes get redistributed throughout the full energy loss range, appearing as weak satellites to each harmonic of the strongly coupled high energy mode.

{}

The inverse case of a more strongly coupled low energy mode with a weakly coupled high energy mode is presented in Fig.~\ref{fig:low}.  In each panel, the high energy mode has a coupling strength of $M_2=50$ meV ($2M_2/\gamma=0.25$) while the low energy mode is either weakly ($2M_1/\gamma=0.5$, $M_1/M_2=2$; Fig.~\ref{fig:low}a) or moderately ($2M_1/\gamma=1$, $M_1/M_2=4$; Fig.~\ref{fig:low}b) coupled.  Similarly to the previous case, the intensity at the first harmonic of the weakly coupled high energy mode is reduced in the two mode model with respect to the independent mode approximation.  In the more extreme case of $M_1/M_2=4$ (Fig.\ref{fig:low}b), the contribution of the high energy mode is nearly lost and the two mode model is very close to the single mode model for only the low energy mode (dashed blue curve), though it deviates strongly from the summation of the two independent modes (red curve).  This is highlighted by the corresponding difference plot beneath.
When the mode couplings are more similar (Fig.~\ref{fig:low}a) the overall spectral shape is rather different between the two mode model and the independent summation.  The two model again shows significantly reduced intensity at the first harmonic of the high energy mode, but this intensity is now clearly redistributed in the tail region.  Consequently, the two mode model deviates significantly both from the sum of the single modes (red curve), and from only the contribution of the low energy mode (dashed blue curve).

While the effect is similar to that observed in Fig.~\ref{fig:high} --  namely, the spectral weight of the weakly coupled mode gets redistributed throughout the spectrum as satellites to each harmonic of the strongly coupled mode -- the impact on the spectra is less obvious to observe in this case.  The difficulty occurs because the first satellite associated with the high energy mode will occur at $\omega_1+\omega_2$, which in this example is a higher energy than the third harmonic of $\omega_1$.  This is demonstrated in the difference plots, which do not show a positive contribution until $\omega_1+\omega_2=105$ meV.  Since the presence of the high energy phonon is no longer obvious in the two mode model (black curve), given experimental data of this form, one might attempt a single mode fit.  Such a fit could yield a greatly exaggerated coupling constant for the low energy mode.

The spectra in Figs.~\ref{fig:equal}-\ref{fig:low} are scaled so that the two mode model and single mode summation give equal intensity for the highest peak in each loss profile.  In general, the intensities of the first harmonic contribution of each mode, {\it i.e.}~for the final states $\ket{n_1,n_2}=\ket{1,0}$ or $\ket{0,1}$, are affected by the occupancy of the other mode in the intermediate state.  This is not necessarily obvious since the model Hamiltonian we consider contains only harmonic oscillators and does not explicitly contain a term for scattering between the phonon modes.  Nevertheless, the two harmonic vibrational modes interact with each other indirectly through the electronic level.  These interactions are significant in the RIXS intermediate state for which it is necessary to include states with high phonon occupancies.  Interference effects of mode mixing in the intermediate state can both increase or decrease the intensity of peaks relative to the independent oscillator case.  Mathematically, this is because the functions $B$ and $D$ are not necessarily positive.  Thus, the intensity of the first harmonic (and all higher harmonics) of each oscillator encodes information about the coupling strength of both modes and one cannot simply assign a single coupling constant to a given peak or series of harmonics.  This challenges the use of detuning as a viable method to extract the coupling strength.  We discuss this point further in section \ref{sec:detuning}.

\section{Distorted Harmonic oscillator}
\label{sec:distorted}

The canonical transformation and related analysis that led to Eq.~1 for the intensity amplitudes of the RIXS phonon features assumed that the vibrational potential energy surface undergoes only a shift of equilibrium position between the ground- and electronic excited-states.  That is, the curvature of the potential energy surface, and consequently the vibrational frequency, remains unchanged.  This is equivalent to the displaced harmonic oscillator model.  In this section, we consider the impact of allowing the curvature of the potential energy surface to change between the ground- and excited-states -- the displaced and distorted harmonic oscillator.  The displaced and distorted oscillator model was previously studied in the context of resonance Raman scattering \cite{Islampour.1995}.

It is clear that the ground- and excited-state vibrational frequencies can differ for molecules or other non-periodic systems.  The same effect may also occur in periodic systems when the RIXS intermediate state induces local vibrational modes.  These excited-state local modes, which may arise due to a change in local symmetry or the perturbation caused by the core-hole potential, can have frequencies appreciably different from the ground state phonon frequencies.  The RIXS intermediate-state at the Ti L-edge of SrTiO$_3$ consists of a $d^1$ configuration.  This induces a local $E$-$e$ Jahn-Teller distortion of the potential energy surface with respect to the $d^0$ ground-state configuration.  The frequency of the excited-state Jahn-Teller modes, 33.5 meV, differ significantly from the corresponding LO-2 mode around 55 meV \cite{Tinte.2008}.  These considerations, coupled with first-principles calculations, have successfully explained the X-ray absorption vibrational linewidths in SrTiO$_{3}$ \cite{Gilmore.2010}.  In the more extreme case of diamond, the $\sigma^*$ excited-state actually hops between two excited-state potential energy surfaces as the local bond length undergoes a change of 15 $\%$ \cite{Ma-diamond.1993}.  Given that the use of RIXS to quantify vibronic coupling strengths is typically applied to strongly correlated materials with localized electrons we expect that it is not uncommon for the RIXS intermediate-state vibrational frequency to differ from the ground-state frequency.  In this section, we investigate the implications of this scenario for the standard single mode Franck-Condon model.

We describe a single Einstein mode coupled to a single electronic excitation with the following Hamiltonian
~
\begin{eqnarray}
    H &=& (1-d^+d) \left [ \omega_0 \, b^+b \right ] \\ \nonumber
    &+& (d^+d) \left [ \epsilon_0 + \tilde{\omega}_0 \, \tilde{b}^{+} \tilde{b} + \sqrt{g} \, \tilde{\omega}_0 \, (\tilde{b} + \tilde{b}^+) \right ] \, .
\end{eqnarray}

\noindent The first term on the right hand side describes the electronic ground-state ($d^{+}d=0$) for which we consider only the ground-state vibrational energy $\omega_0$.  The second term is for the electronic excited-state ($d^{+}d=1$) and contains the energy of the excited electronic level $\epsilon_0$ referenced with respect to the electronic ground-state, the vibrational energy $\tilde{\omega}_0$ in the excited-state, and a coupling term between the electronic level and the vibrational mode.  Tildes indicate excited-state quantities so that $\tilde{b}^{+}$ ($\tilde{b}$) creates (destroys) a quantum of oscillation of energy $\tilde{\omega}_0$ on the excited-state potential energy surface.

This problem is equivalent to the displaced and distorted harmonic oscillator.  To derive the RIXS amplitudes for this model we first address the displaced aspect of the oscillator in a similar way to how it was treated for the undistorted case.  That is, by applying the canonical transformation with a generating function $\tilde{S} = \sqrt{g} \tilde{\omega}_0 d^+d(\tilde{b}^{+}-\tilde{b})$ that depends on the excited-state quantities.  With this, the vibrational part of the RIXS amplitudes become
~
\begin{equation}
    \label{7}
     A_{n}= \sum_{\tilde{m}} \frac{\bra{n} e^{-\tilde{S}}\ket{\tilde{m}}\bra{\tilde{m}}e^{\tilde{S}}\ket{0}}{z-\tilde{\omega}_0(\tilde{m}-g)  } \ .
\end{equation}

\noindent However, this expression differs from Eq.~\eqref{eq:simple.FC} because the intermediate-state vibrational wavefunctions $\ket{\tilde{m}}$ still correspond to the relatively distorted excited-state potential energy surface.  To account for the distorted aspect of the oscillator, we now define a second transformation between the ground- and excited-state vibrational bases
~
\begin{equation}
    \label{8}
    \ket{n}=\sum_{\tilde{n}}\ket{\tilde{n}} \braket{\tilde{n}|n}=\sum_{\tilde{n}} X_{\tilde{n},n}(\beta)\ket{\tilde{n}} \, .
\end{equation}

\noindent The analytical expression for the transformation matrix $X_{\tilde{n},n}(\beta)$ can be found using second quantization approaches \cite{Nishikawa1977,Palma,Witschel_1973} or by exactly evaluating overlap integrals for two distorted harmonic  potentials\cite{Chang,TS_2016} in terms of Hermite polynomials $H_j$.  This gives
~
\begin{widetext}
\begin{equation}
    \label{9} 
    X_{\tilde{n},n}(\beta)=\sqrt{\frac{1}{2^{n+\tilde{n}}n!\tilde{n}!}}\sqrt{\frac{2\beta}{1+\beta^2}} 
    \sum_{k=0}^{n} \sum_{\tilde{k}=0}^{\tilde{n}}
    \binom{n}{k} \binom{\tilde{n}}{\tilde{k}} 2^{k+\tilde{k}} \beta^{\tilde{k}} H_{n-k}(0) H_{\tilde{n}-\tilde{k}}(0) I(k+\tilde{k}) \ 
\end{equation}

\noindent where the dimensionless parameter $\beta=\sqrt{\tilde{\omega}_0/\omega_0}$ reflects the change in the phonon energy  and the function I(K) is  
~
\begin{equation}
I(K) = \left \{ 
\begin{matrix}
0 \hspace{0.8in} \text{if K is odd} \\
\frac{(K-1)!!}{\sqrt{1+\beta^2}^K} \hspace{0.4in} \text{if K is even} \\
\end{matrix}
\right . \, \, .
\end{equation}

\noindent The final RIXS amplitudes for the distorted oscillator are
~
\begin{equation}
    \label{3a}
    A_{n}=\sum_{m,l,k} X_{n,l}(\beta) X_{k,0}(\beta)
    \frac{B_{l,m}(g) \, B_{m,k}(g)}{z-\tilde{\omega}_0(m-g) } \ .
\end{equation}
\end{widetext}

\noindent Eq.~\eqref{3a} gives the RIXS amplitudes corresponding to the basic Holstein Hamiltonian generalized to the case of a change in the curvature of the potential energy surface between the ground and excited electronic states. 

\begin{figure}
    \centering
    \includegraphics[width=1.\linewidth]{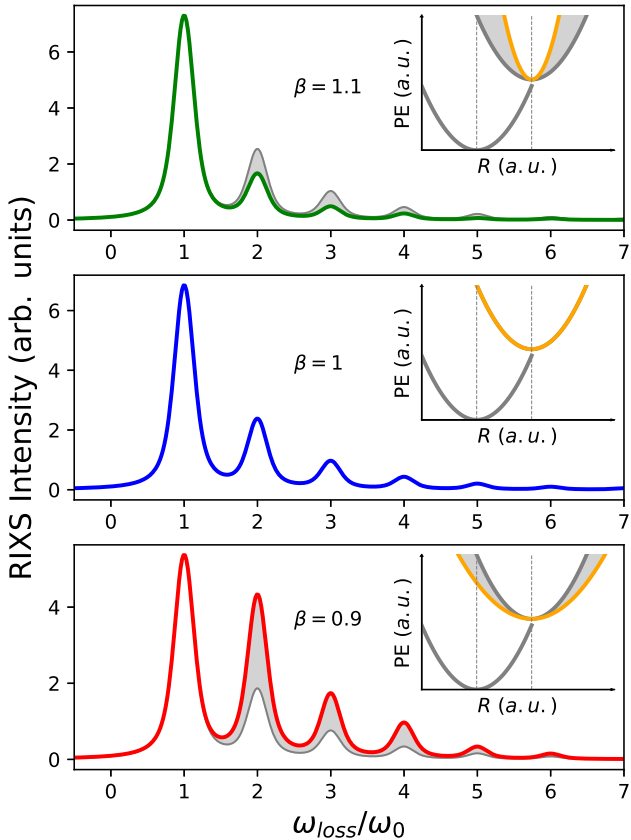}
    \caption{Influence of distortion of the curvature of the excited-state potential energy surface (PES) on RIXS vibrational intensities.  Insets show schematics of the relative curvatures of the ground-state and excited-state PES.  The actual excited-state PES is indicated in yellow and compared to the ground-state surface in grey.  Main figures present the corresponding RIXS vibrational intensities according to Eq.~\eqref{3a}. All calculations were done for a coupling constant $M=150$ meV, inverse core-hole lifetime $\gamma/2=200$ meV and ground state phonon energy $\omega_{0}=100 $ meV. Grey curves represent spectra for the case of no distortion.} 
    \label{fig:my_label}
\end{figure}{}

In Fig.~\ref{fig:my_label} we consider three scenarios: $\tilde{\omega}_0<\omega_0$ ($\beta<1$), $\tilde{\omega}_0=\omega_0$ ($\beta=1$), and $\tilde{\omega}_0>\omega_0$ ($\beta>1$).  The change in curvature of the excited-state potential energy surface makes a significant impact on the relative intensities of the RIXS phonon features.  The relative intensities of higher order harmonics decrease for $\beta>1$ and increase for $\beta<1$.  The effect is significant for a 10 $\%$ deviation of $\beta$ ($\sim$20 $\%$ change the oscillator energy).  This can cause a misquantification of the coupling strength if the change of the potential energy surface is not accounted for.  Unfortunately, to accurately extract quantitative information from the phonon features one must know the curvature of the excited-state potential energy surface with respect to its ground-state counterpart.  It is possible that this could be estimated experimentally by analyzing the linewidths of x-ray absorption spectra or the vibrational spectra associated with local modes around dopants, but this would be challenging in general.

\section{Impact on detuning}
\label{sec:detuning}

The single mode Franck-Condon model, Eq.~\eqref{eq:simple.FC}, is most often used to quantify the vibronoic coupling strength by fitting the relative intensities of a harmonic series in the RIXS loss profile.  However, an alternative approach has recently been suggested in which only the intensity of the first harmonic is required \cite{Rossi.2019}.  In this case, Eq.~\eqref{eq:simple.FC} is used to fit the reduction in intensity of the first harmonic with respect to the detuning of the incident photon energy from the resonance energy.  The authors applied this to the Cu-O bond stretching mode of NdBa$_2$Cu$_3$O$_6$, and later simultaneously to both the bond stretching and buckling modes of the same material \cite{Braicovich.2019}.  Separately, this scheme was recently used to distinguish the contributions from the zone center and zone boundary modes in graphite \cite{YiDe.2019}.  However, the impact of a second mode on the detuning behavior is unknown.  We begin this section by investigating that question.  We subsequently probe the detuning behavior of the single-mode displaced and distorted oscillator introduced in the previous section.

\begin{figure}
\includegraphics[width=1.0\linewidth,angle=0]{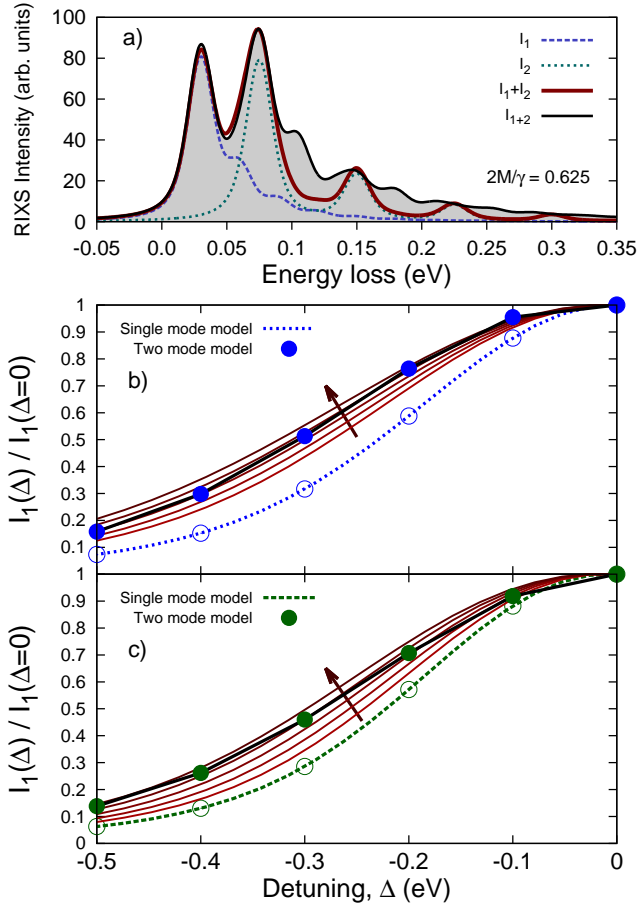}
    \caption{(a) RIXS loss profile for two phonon modes at zero detuning with $\omega_1=30$ meV, $\omega_2=75$ meV, and $M_1=M_2=125$ meV.  The color scheme is the same as used in Fig.~\ref{fig:equal}. (b) The relative intensity of the first harmonic of the low energy mode as a function of detuning.  The detuning curve from the two mode model (black line with blue symbols) deviates strongly from that for the independent mode approximation (dashed blue curve).  Red background curves are generated from the single mode model for coupling strengths between 160-200 meV in steps of 10 meV, increasing in the direction of the arrow.  (c) Same as the middle panel, but for the high energy mode.  Red background curves are generated from the single mode model for coupling strengths between 140-180 meV in steps of 10 meV.}
    \label{fig:detune-eq}
\end{figure}

We again consider a low energy mode with $\omega_1=30$ meV and a high energy mode with $\omega_2=75$ meV while keeping the core-hole lifetime fixed at $\gamma/2=200$ meV.  We treat first the case that the two modes have equal coupling strength of $M_1=M_2=125$ meV ($2M/\gamma=0.625$); the results are presented in Fig.~\ref{fig:detune-eq}.  The top panel contains the total RIXS loss profile at zero detuning while the middle panel gives the variation of the peak intensity with the detuning at 30 meV ($\omega_1$) and the bottom panel shows the same for the 75 meV peak ($\omega_2$).  

The detuning curves obtained from the two mode model deviate significantly from those given by the independent mode approximation (dashed curves within each panel).  Furthermore, the curves produced by the two mode model cannot be reproduced by any value of the coupling strength without changing the core-hole lifetime.  In particular, for small detuning the low (high) energy mode initially follows a single mode detuning curve for a coupling strength of 220 meV (150 meV), but for larger detuning the low (high) energy curve tracks the detuning curve for 180 meV (180 meV).  The actual coupling strength for both modes is 125 meV, revealing an error of approximately 50 $\%$ that depends on the detuning value and mode.

It is interesting to note the opposite behavior of the low and high energy modes.  The effective coupling strength of the low energy mode deviates most strongly from the true coupling strength at small detuning and slowly approaches the correct coupling strength for large detuning.  The opposite trend is observed for the higher energy mode.  This is exactly what was observed (though not discussed) during the simultaneous study of the breathing (70 meV) and buckling (30 meV) modes of NdBa$_2$Cu$_3$O$_6$ \cite{Braicovich.2019}.  In that work, the intensity of the first harmonic of the higher energy breathing mode versus detuning clearly traverses several calculated single-mode detuning curves starting from around $g=2$ ($M=100$ meV) at low detuning and surpassing $g=8$ ($M=200$ meV) at higher detuning.  The authors report the extracted coupling strength as the single value $g=4$ ($M=130$ meV).  The results in Fig.~\ref{fig:detune-eq}c suggest a more likely value is $M<100$ meV.

{}

\begin{figure}
\includegraphics[width=1.0\linewidth,angle=0]{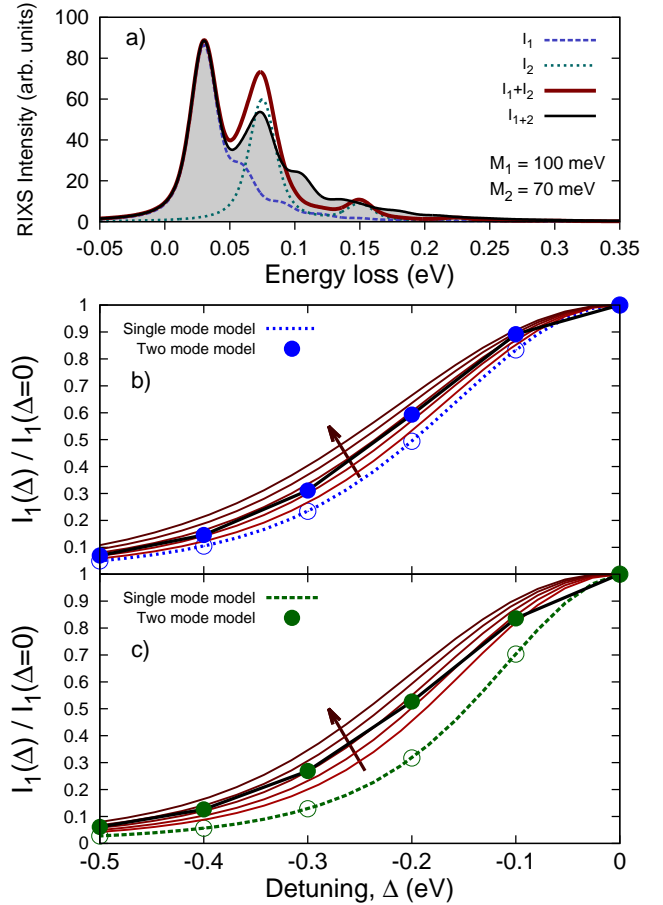}
    \caption{(a) RIXS loss profile for two phonon modes at zero detuning with $\omega_1=30$ meV, $\omega_2=75$ meV, $M_1=100$ meV and $M_2=70$ meV.  Detuning curves for the low energy mode (b) and the high energy mode (c).  Presentation style of all panels is the same as in Fig~\ref{fig:detune-eq}.  Background single mode detuning curves represent coupling strengths between 110-150 meV (b) and 100-140 meV (c) in steps of 10 meV.}
    \label{fig:detune-lw}
\end{figure}{}

When one mode is coupled more strongly than the other we find that the error in the inferred coupling strength of the more strongly coupled mode is reduced while that for the more weakly coupled mode increases.  This case is considered in Fig.~\ref{fig:detune-lw} for which the actual coupling strengths are $M_1=100$ meV for the low energy mode and $M_2=70$ meV for the high energy mode.  The detuning curve of the low energy peak follows the $M=140$ meV curve at low detuning and approaches the $M=120$ meV curve by a detuning of 0.5 eV.  This represents a 20-40 $\%$ error for the more strongly coupled mode.  On the other hand, the high energy mode initially follows the $M=100$ meV detuning curve, but shifts to the $M=130$ meV curve by 0.5 eV detuning.  This gives an error spanning the range of 40-85 $\%$.  For the opposite case of a more strongly coupled high energy mode (not shown), the results follow similar trends.  That is, the error in inferred detuning of the more strongly coupled mode is approximately 30 $\%$ while that for the more weakly coupled mode exceeds 100 $\%$ at low detuning and approaches a 70 $\%$ error by a detuning of 0.5 eV.

Most systems of interest will exhibit at least two RIXS-active modes.  When these have similar coupling strengths, the use of detuning will yield a value for the coupling strength that is exaggerated, easily on the order of 50 $\%$.  If one mode clearly dominates the others in terms of coupling strength, the use of detuning to quantify the coupling strength becomes more reliable for the strongly coupled mode, but is very unreliable for the more weakly coupled modes with errors exceeding even 100 $\%$.  It appears that in all cases, the coupling strength inferred from detuning will overestimate the true value to some degree when more than one mode is active.

\begin{figure}
    \centering
    \includegraphics[width=0.7\linewidth,angle=270]{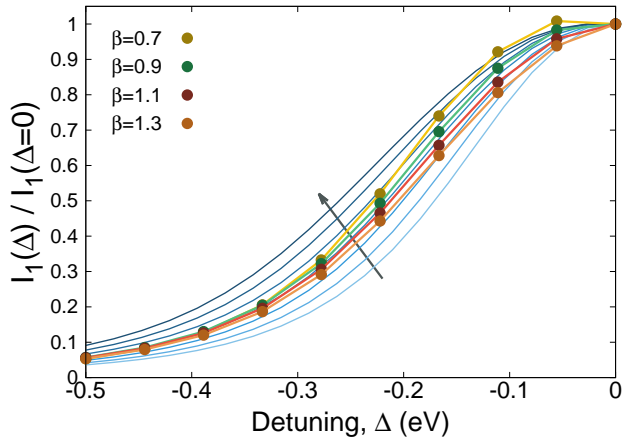}
    \caption{Intensity of the first phonon harmonic versus detuning for the displaced-distorted oscillator {using $\omega_0=100$ meV, $\gamma/2=200$ meV and $M=125$ meV.}  The detuning curves for $\beta=0.7$ (yellow), $\beta=0.9$ (green), $\beta=1.1$ (red), and $\beta=1.3$ (orange) are compared to undistorted single oscillator detuning curves for coupling strengths between 95-155 meV in steps of 10 meV, increasing with the arrow.}
    \label{fig:distort-detune}
\end{figure}

The detuning curves for the displaced-distorted oscillator model are presented in Fig.~\ref{fig:distort-detune} for the parameter set $\omega_0=100$ meV, $M=125$ meV, $\gamma/2=200$ meV with $\beta \in \{0.7,0.9,1.1,1.3\}$.  For $\beta<1$, the detuned value follows an inflated value of the coupling constant for small detunings while for $\beta>1$ the intensity of the first harmonic suggests a weaker effective coupling strength at small detuning.  The detuning trends for both $\beta<1$ and $\beta>1$ gradually converge to the correct detuning curve for large detuning.

{}

At the smallest value of $\beta$ we observe that the detuning curve is slightly non-monotonic; the intensity for small detuning is actually higher than the zero detuning value.  While this behavior may seem non-physical, it has been observed experimentally in the detuning curve for the $\sigma^*$ excitation in graphite \cite{YiDe.2019, YiDe.private}.  A difference in curvature between the ground- and excited-state potential energy surfaces offers a possible explanation of this observation.

\begin{figure}
    \centering
    \includegraphics[width=1\linewidth,angle=0]{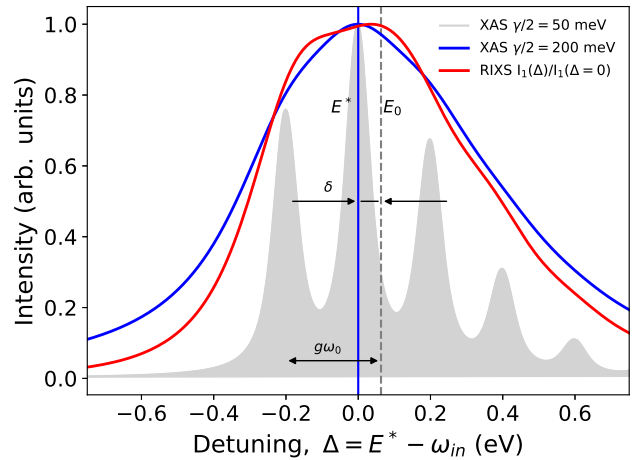}
    \caption{Intensity of the X-ray absorption spectrum (blue) and first phonon harmonic (red) versus detuning for the single mode displaced oscillator using $\omega_0=200$ meV, $\gamma/2=200$ meV and $M=230$ meV.  The grey profile reveals the vibronic features in the X-ray absorption for an artificially long core-hole life-time ($\gamma/2=50$ meV). The zero of detuning is set to the maximum of XAS, $E^*$, which differs by $\delta$ from the energy of non-interacting quasiparticle $E_0$. }
    \label{fig:detune-schematic}
\end{figure}

An alternative explanation for this unusual detuning behavior in graphite is that the standard Franck-Condon model can produce this type of non-monotonic detuning curve in the case of simultaneous strong coupling and long core-hole lifetime.  This introduces a delicate issue about the application of detuning studies.  Experimentally, zero detuning is defined with respect to the maximum of the given feature in the X-ray absorption spectrum (XAS).  However, in the case of strong coupling and large phonon energies the peak of the XAS can be offset from the actual electronic resonance.  We illustrate this in Fig.~\ref{fig:detune-schematic}.  A finely-resolved model X-ray absorption spectrum, including vibrational excitations, is exhibited by the grey shaded region.  The true energy position of the electronic level is $E_0$, however, due to the vibronic coupling, the first eigenvalue of the spectrum is downshifted in energy by $g\omega_0$ where $\omega_0$ is the phonon energy.  Broadening the spectrum to account for the core-hole lifetime gives the blue curve for the XAS.  The peak of the XAS is shifted slightly, by $\delta$, from the energy of the electronic level $E_0$.  The shift $\delta$ depends on the coupling strength, phonon energy and core-hole lifetime.  An experimental detuning curve assumes that zero detuning coincides with $\omega_i = E_0 - \delta$ while the calculated detuning curve sets the zero of detuning as $\omega_i = E_0$.  This offset could lead to a minor increase in the intensity of the first phonon harmonic of the RIXS loss profile for small detunings.  In such cases, measuring the detuning curve both above and below resonance may offer some clarity.

{}

\section{Conclusion}

We have considered two generalizations of the standard Franck-Condon description of the phonon contribution to resonant inelastic X-ray scattering.  In the case that a second Einstein oscillator is added to the Holstein Hamiltonian the correct RIXS signal, obtained through an exact analytical solution of the two mode model, differs significantly from the signal obtained by independently summing the contributions of two single oscillator models.  Not only do new phonon peaks appear at energies corresponding to combinations of the two oscillator energies, but the relative intensities of all peaks differ with respect to the intensities obtained from the independent mode approximation.  This occurs because, while the oscillators do not couple directly, they are indirectly coupled through the electronic level.  Even when the final state contains only a single quantum of a single mode, the intermediate states contributing to its peak intensity can include highly occupied levels of both oscillators.  The consequence is that peak intensities reflect a complicated combination of the coupling strengths of all active modes.  The overall result may be viewed approximately as a convolution of the {\it amplitudes} of two independent mode signals with the caveat that the total spectral weight of the phonon loss features is generally greater in the two mode model compared to the independent mode approximation.  Since for most crystalline materials multiple phonon modes couple to the electronically excited state the mode mixing effect should not be ignored.

Even when only a single mode is active, this mode still has an associated wavevector and one expects to observe a dispersion not only of the mode energy, but also of the vibronic coupling strength.  Thus far, all studies of the q-dependence of the coupling strength as quantified by RIXS assume that each q-point may be analysed independently.  Our two mode model demonstrates that the RIXS intensities at each q-point are not independent, even at the first harmonic.  For example, when the momentum transfer is near the $\Gamma$-point, the first phonon harmonic, normally thought to be purely associated with the $q=0$ phonon, will be impacted through the intermediate state by the coupling strength of $\pm q$ pairs throughout the Brillouin zone (in fact, higher-order combinations may also contribute).  This effect is non-negligible since phonon occupancies can reach significant oscillator levels in the intermediate state.  

We also investigated the impact on phonon peak intensities in RIXS spectra when the oscillator frequency differs between the ground and excited states.  The effect of a change in the vibrational frequency is similar to that obtained by changing the coupling strength and/or core-hole lifetime while holding the vibrational frequency constant.  It is therefore difficult to distinguish the difference of the two effects experimentally.  Without knowledge of the excited-state vibrational frequency, use of the relative intensities of a harmonic series to quantify the vibronic coupling strength can become ambiguous.

Throughout this work we have assumed harmonic oscillators linearly coupled to an electronic state.  While this is typically a good approximation for the RIXS initial and final states of periodic systems, the RIXS intermediate state can reach high oscillator levels.  This suggests that the impact of anharmonic effects on the evolution of the intermediate state should be probed in future work.

We find that phonon generation during a RIXS measurement is a complex process that cannot be adequately described quantitatively by simple models in general.  This highlights the need for the comparison of experimental results to unbiased, first-principles calculations of the phonon contribution to RIXS in order to further advance our understanding of this new measurement technique.  It will be advantageous in future work to further develop models of the phonon contribution to RIXS based on the momentum-dependent Fr\"{o}hlich Hamiltonian beyond what has already been presented \cite{Devereaux.Frohlich}.

\section{Acknowledgements}

We thank J.~Pelliciari and T.~Ziman for valuable discussions.  KG was supported by the U.S. Department of Energy, Office of Science, Basic Energy Sciences as part of the Computational Materials Science Program.

\bibliographystyle{natbib}




\end{document}